\documentclass[a4paper]{jpconf}
\usepackage{amsmath,amssymb}
\usepackage{graphicx}
\usepackage{color}

\begin{document}
\title{Quantum Monte Carlo study of the transverse-field Ising model on a frustrated checkerboard lattice}

\author{
H. Ishizuka, Y. Motome, N. Furukawa$^{A,B}$, and S. Suzuki$^A$
}

\address{
Department of Applied Physics, University of Tokyo, Japan \\
$^A$Department of Physics and Mathematics, Aoyama Gakuin University, Japan \\
$^B$Multiferroics Project, ERATO, Japan Science and Technology Agency (JST),\\
c/o Department of Applied Physics, University of Tokyo, 
Japan
}

\ead{ishizuka@aion.t.u-tokyo.ac.jp}

\begin{abstract}
We present the numerical results for low temperature behavior of the transverse-field Ising model on a frustrated
checkerboard lattice, with focus on the effect of both quantum and thermal fluctuations.
Applying the recently-developed continuous-time quantum Monte Carlo algorithm, we compute the magnetization and susceptibility down 
to extremely low temperatures while changing the magnitude of both transverse and longitudinal magnetic fields.
Several characteristic behaviors are observed, which were not inferred 
from the previously-studied quantum order from disorder at zero temperature, 
such as a horizontal-type stripe ordering at a substantial longitudinal field and 
a persistent critical behavior down to low temperature in a weak longitudinal field region.
\end{abstract}

\section{Introduction}

Frustration due to the geometrical structure of systems has been one of the major topics in condensed matter physics.
The interest covers a wide range of fields, such as frustrated magnetism~\cite{Diep2004}, 
proton ordering in hydrogen-bonded systems~\cite{Pauling1933,Slater1941}, and 
anomalous transport and multiferroics in transition metal oxides~\cite{Anderson1956,Taguchi2001,Kimura2003}.
The geometrical frustration often prevents the system from selecting a unique ground state, 
instead gives rise to classically degenerate ground states; i.e., 
a macroscopic number of different configurations of the system variables 
lead to the same ground state energy at the classical level. 
As a result, the ground state remains to be disordered and bears finite residual entropy.
Such macroscopic degeneracy provides a fertile ground for various peculiar phenomena as
it makes the system to be extremely sensitive to perturbations,
such as remnant interactions, external fields, and fluctuations.

Among such degeneracy-lifting mechanisms, quantum and thermal fluctuations have attracted particular
interest. These fluctuations sometimes lift the ground-state degeneracy and induce some particular
ordering, as known by order from disorder~\cite{Villain1980}.  There, the ordering is selected from the
manifold to maximize the associated entropy (zero-point entropy in the case of quantum fluctuation). 
This entropic effect plays a decisive role at low temperatures ($T$) and causes many fascinating phenomena
in the frustrated systems.

One of the minimal models for studying the order-from-disorder phenomenon is the transverse-field Ising
model (TIM). In the absence of geometrical frustration, in general, the model develops a long-range order
at low $T$, while the ordering is suppressed and a quantum paramagnetic state is induced by the transverse
field. When the classical ground state is macroscopically degenerate under strong frustration, 
quantum fluctuations by the transverse field and/or thermal fluctuations by temperature can induce a
particular ordering via the order-from-disorder mechanism. Effect of thermal fluctuations has been studied
mainly in the absence of the transverse field, i.e., for pure Ising models without the transverse field~\cite{Diep2004,Liebmann1986}. 
Meanwhile, effect of quantum fluctuations has also been studied.
For example, it was shown that TIM on a variety of frustrated lattices exhibit several nontrivial
behaviors at $T=0$, such as a bond ordering and Kosterlitz-Thouless transition
~\cite{Moessner2000,Moessner2001}. In general, the thermal and quantum fluctuations do not necessarily lead
to the same effect, and the relation between them is of particular interest to 
explore yet another order-from-disorder phenomenon.

In this contribution, we present our numerical results for the order-from-disorder phenomena in TIM 
in a wide range of temperature and the transverse/longitudinal magnetic fields. We consider the TIM on a
two-dimensional checkerboard lattice and study its low-$T$ physics by a sophisticated quantum Monte Carlo
(QMC) technique.
As a result, in the intermediate longitudinal-field regime, we find instability
toward a Neel order, in accordance with the previous report~\cite{Moessner2001}.
On the other hand, for both the weaker and stronger longitudinal fields, 
our results indicate different behaviors from the previous report.
Under the weak field, featureless magnetic susceptibility is observed down to
extremely low $T$, which implies a very weak proximity effect
to the predicted Neel ground state or a possibility of another state in the
$T=0$ limit.
On the contrary, in the strong field, the system 
indicates instability toward an unexpected horizontal-type stripe ordering.

\section{Model and method}

\begin{figure}[h]
\includegraphics[width=20pc]{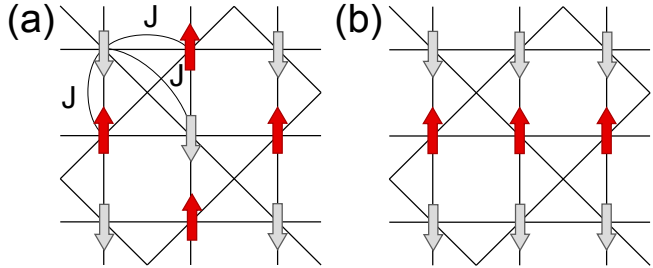}\hspace{2pc}
\begin{minipage}[b]{13pc}
      \caption{
      Schematic pictures of the transverse-field Ising model on the checkerboard lattice.
      (a) Neel type order and (b) horizontal stripe order.
      }
      \label{fig1}
\end{minipage}
\end{figure}

To investigate the effect of quantum and thermal fluctuations on the classically degenerate ground state in
frustrated systems, here we focus on the frustrated checkerboard Ising model with transverse and
longitudinal magnetic fields.  The Hamiltonian is given as:
\begin{eqnarray}
H= J\sum_{\langle i,j\rangle} s_i^z s_j^z + \Gamma \sum_i s_i^x + h \sum_i s_i^z,
\label{eq:H}
\end{eqnarray}
where $s_i^\alpha$ is the Pauli spin operator at site $i$, and $J$ is the Ising interaction
between nearest-neighbor sites $\langle i,j\rangle$ on the checkerboard lattice (see Fig.~\ref{fig1});
$\Gamma$ and $h$ are the transverse and longitudinal fields, respectively. In the following, we consider
the antiferromagnetic case, $J>0$, and set the energy scale by $J=1$.

When $\Gamma=h=0$, the model remains to be disordered down to zero $T$~\cite{Liebmann1986}. The lowest
energy is achieved by enforcing a simple local constraint --- a two-up two-down local spin configuration 
for all the plaquettes with crisscrossing interaction, similar to the so-called ice rule in water ice
~\cite{Pauling1933}. Consequently, a macroscopic number of different spin configurations give the same
lowest energy: The ground state is macroscopically degenerate and the residual entropy is estimated to be
$\sim \frac{N}{2} \log \frac32$ ($N$ is the number of spins). The spin correlation, however, is critical in
the sense that it decays algebraically as a function of distance. The situation is unaltered as $h$
increases up to $h=2J$ at $\Gamma=0$. 
(For larger $h$, the ground state consists of ``three-up one-down" configurations, 
which is also macroscopically degenerate.)
When the quantum fluctuation sets in by switching on $\Gamma$, it was predicted, from perturbative
considerations in $\Gamma/h$, that a Neel 
order is induced in the $0<h<2J$ 
region at $T=0$~\cite{Moessner2001} [see Fig.~\ref{fig1}(a)]. 
Hereafter, we focus on the $0<h<2J$ region.

To investigate the low-$T$ behaviors of this model with taking account of both quantum and thermal
fluctuations, we conducted numerical calculations by QMC method based on the path-integral
approach~\cite{Suzuki1976_1,Suzuki1976_2}. Here, we employed the recently-developed continuous-time
algorithm~\cite{Nakamura2008} to approach extremely low $T$. At low $T$, the MC sampling suffers from the
slow relaxation due to the severe frustration. To accelerate the relaxation process, we applied the
loop-flip algorithm developed for frustrated Ising-type spin systems~\cite{Rahman1972,Melko2001}. In our
calculations, the loops are formed in the real space at a particular imaginary-time slice, and flipped
together with all the spin variables along the imaginary-time direction, i.e., on the torus defined by the
loop in space and time. In addition, the replica exchange method was used to further suppress the slowing
down~\cite{Hukushima1996}. 
For the present system, the replicas are exchanged along a constant-$\Gamma/T$ line because the Boltzmann weight
depends on the number of domain walls in the imaginary-time direction which is proportional to $\Gamma/T$. 

Calculations were done for $N=4\times L^2$ site systems with $L=12$ under the periodic boundary conditions. 
Observables were computed typically for 120000 samplings after 30000 steps of thermalization.
The results were divided into eight bins to evaluate the statistical errors.

\section{Results and discussions}

\begin{figure}[h]
\begin{minipage}{38pc}
\includegraphics[width=38pc]{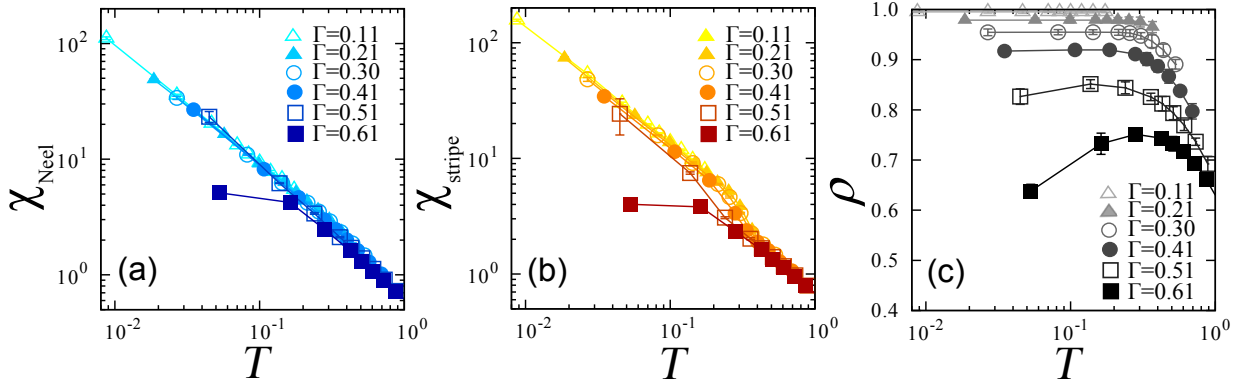}
      \caption{
      $T$ dependence of (a) $\chi_{\rm Neel}$, (b) $\chi_{\rm stripe}$, and (c) local correlation parameter $\rho$ at
      $h=0.4$ for the transverse field ranging from $\Gamma=0.11$ to $0.61$. 
      }
      \label{fig2}
\end{minipage}
\end{figure}

Figure~\ref{fig2} shows the QMC results for the susceptibility at a relatively weak longitudinal field
$h=0.4$ with varying $\Gamma$. Figure~\ref{fig2}(a) presents the susceptibility for the Neel-type order
parameter [Fig.~\ref{fig1}(a)], 
\begin{equation}
\chi_{\rm Neel} = \frac{N}{T} ( \langle m_{\rm Neel}^2 \rangle - \langle | m_{\rm Neel} | \rangle^2 ) \ ; \quad 
m_{\rm Neel} = \frac{1}{N} \sum_i s_i^z (-1)^{i_x+i_y}, 
\end{equation}
while Fig.~\ref{fig2}(b) is that for the horizontal-type stripe order parameter [Fig.~\ref{fig1}(b)], 
which we consider as another candidate of order from disorder,
\begin{equation}
\chi_{\rm stripe} = \frac{N}{T} ( \langle m_{\rm stripe}^2 \rangle - \langle m_{\rm stripe}\rangle^2 ) \ ; \quad
m_{\rm stripe}^2 = \Big(\frac{1}{N} \sum_i s_i^z (-1)^{i_x} \Big)^2 + 
\Big(\frac{1}{N} \sum_i s_i^z (-1)^{i_y} \Big)^2. 
\end{equation}
Note that this stripe state belongs to the two-up two-down manifold and 
is different from the diagonal one discussed in the three-up one-down manifold for $2J<h<6J$~\cite{Moessner2001}.
Figure~\ref{fig2}(c) shows the local correlation parameter given by
$\rho = \frac{2}{N}\sum_p f(p)$, where the sum runs over all plaquettes. 
$f(p)$ is a function defined for each plaquette $p$ as 
$f(p)= 1$ for $\sum_{i \in p} s_i^z = 0$ and otherwise $f(p)=-3/5$; 
$\rho \to 1$ when all the plaquettes satisfy the two-up two-down spin configuration, 
and $\rho \to 0$ when the spins are completely disordered, 
i.e., as $T \to \infty$ and/or $\Gamma \to \infty$. 
As shown in Figs.~\ref{fig2}(a) and \ref{fig2}(b), for weak $\Gamma \lesssim 0.5$, both the susceptibilities increase with
decreasing $T$ in Curie-law like behavior, i.e., $\chi \propto T^{-1}$. A shoulder-like feature observed at $T \sim 0.3$ in
$\chi_{\rm stripe}$ corresponds to a formation of the ice-rule type local configurations,
as indicated by the saturation of $\rho$ in Fig.~\ref{fig2}(c).
These behaviors suggest that the system is in the correlated regime 
with satisfying the local constraint, but remains critical, 
without choosing either Neel or stripe ordering, presumably because of the strong frustration.
We also calculate the momentum dependence of the susceptibility, and 
confirm that it remains featureless without showing any peak in the momentum space.
It is surprising that the critical behavior is robustly observed down to very low $T \sim 0.01$. 
This implies that the order-from-disorder mechanism is extremely weak or ineffective in this region.
On the other hand, the results for $\Gamma \gtrsim 0.6$ show saturating behavior at low $T$.
In this region, $\Gamma$ is strong enough to disturb the ice-rule type configurations [Fig.~\ref{fig2}(c)],
and the ground state turns into a quantum paramagnetic state.

\begin{figure}[h]
\begin{minipage}{38pc}
      \includegraphics[width=38pc]{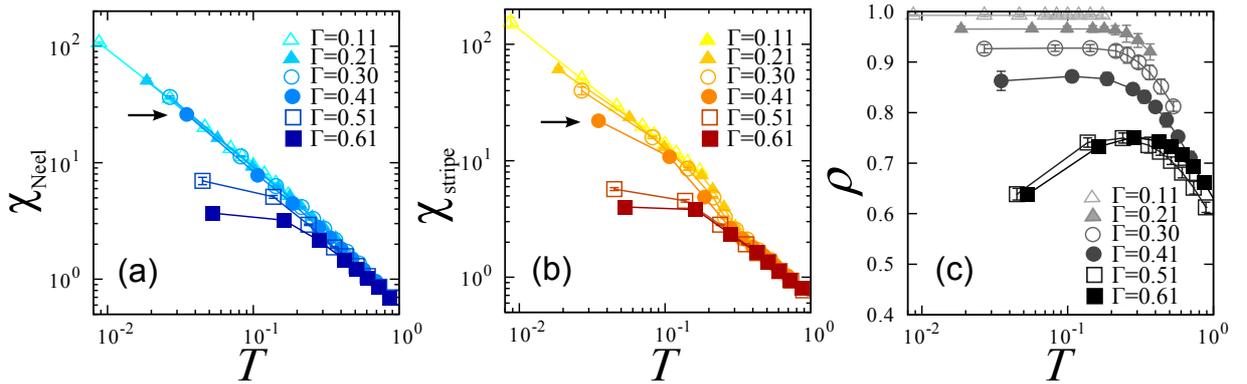}
      \caption{
      $T$ dependence of (a) $\chi_{\rm Neel}$, (b) $\chi_{\rm stripe}$, and (c) local correlation parameter $\rho$
      at $h=0.9$ for the transverse field ranging from $\Gamma=0.11$ to $0.61$.
      The arrows in (a) and (b) indicate a
      contrasting behavior between $\chi_{\rm Neel}$ and $\chi_{\rm stripe}$ in the intermediate $\Gamma$ region.
      }
      \label{fig3}
\end{minipage}
\end{figure}

With increasing the longitudinal field, a contrasting behavior between the two susceptibilities shows up in
the vicinity of phase transition to the quantum paramagnetic state. Figure~\ref{fig3} shows the results at
$h=0.9$. In the region $\Gamma \lesssim 0.3$, both $\chi_{\rm Neel}$ and $\chi_{\rm stripe}$ diverge toward
$T=0$, in a similar manner to the results for $\Gamma \lesssim 0.5$ in Fig.~\ref{fig2}. On the other hand,
before entering the quantum paramagnetic regime for $\Gamma \gtrsim 0.5$ where both $\chi_{\rm Neel}$ and
$\chi_{\rm stripe}$ saturate, a qualitatively different behavior is observed: At $\Gamma \sim 0.40$,
$\chi_{\rm stripe}$ begins to deviate from the Curie-law like behavior and tends to saturate, while
$\chi_{\rm Neel}$ remains to diverge (indicated by the arrows in Fig.~\ref{fig3}). This implies that the
stripe-type fluctuation is suppressed and the Neel-type ordering is likely favored in the low-$T$ limit.
Similar behavior is observed in a finite region on the verge of the quantum paramagnetic phase for
$0.5 < h < 1.0$ (see the phase diagram in Fig.~\ref{fig5} below).

\begin{figure}[h]
\begin{minipage}{38pc}
\includegraphics[width=30pc]{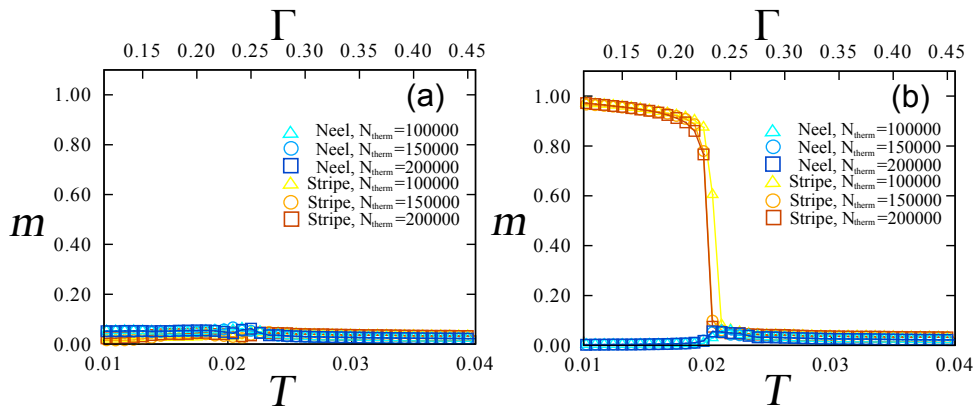}
      \caption{
      $T$ dependence of $m_{\rm Neel}$ and $m_{\rm stripe}$ at $h=1.6$ and $\Gamma/T=\tan(85^\circ)$ 
      for varying thermalization steps $N_{\rm therm}$. (a) is the results starting from the
      perfectly-Neel-ordered configuration and (b) is from the perfectly-horizontal-stripe-ordered one.
      }
      \label{fig4}
\end{minipage}
\end{figure}

For larger longitudinal fields, yet another behavior appears. For $h \gtrsim 1.0$, the MC relaxation
becomes very slow at low $T$ 
even with the use of the loop-flip algorithm and the replica-exchange technique. 
The situation is demonstrated in Fig.~\ref{fig4}. The results show the relaxation of the order parameters
$m_{\rm Neel}$ and $m_{\rm stripe}$ at $h=1.6$ 
along a constant $\Gamma/T = \tan(85^\circ)$ axis; the data are measured for 50000 MC
steps after particular thermalization steps of $N_{\rm therm}$ starting from the initial configuration with
corresponding perfect order: 
Figures~\ref{fig4}(a) and \ref{fig4}(b) show the results starting from
the perfectly-Neel-ordered state and the perfectly-horizontal-stripe-ordered state, respectively.
As shown in Fig.~\ref{fig4}, $m_{\rm Neel} \sim 0$ 
after $N_{\rm therm} = 10^5$ thermalization independent of the initial state.
On the other hand, the $m_{\rm stripe}$ shows 
strong dependence on the initial state even after $N_{\rm therm} \sim 10^{5}$ steps. 
A possible origin of the freezing behavior is strong first order transition to the horizontal-stripe-ordered state.
Although the extremely slow relaxation prevents us from obtaining numerically-converged results, 
the results suggest that the system has an instability toward the horizontal stripe ordering at low $T$ in this high-$h$ region. 

Summarizing the results above, we deduce a `phase diagram' in Fig.~\ref{fig5}, which indicates the low-$T$
instability anticipated from the MC data. In the strong longitudinal field region,
$1.0 \lesssim h \lesssim 1.8$ and $\Gamma \lesssim 0.2$, our MC results indicate an instability toward the
horizontal-stripe-type ordering below some particular temperature, as exemplified in Fig.~\ref{fig4}(b). Although we
could not directly confirm the phase transition because of the extremely slow MC relaxation, we expect this
phase to appear at a finite temperature. In the intermediate longitudinal field region,
$0.5 \lesssim h \lesssim1.0$ and $0.3 \lesssim \Gamma \lesssim 0.5$, the results appear to favor the
Neel-type instability. The Neel temperature appears to be extremely low, less than $T = 0.01$. In the
remaining low-$h$ region for $\Gamma \lesssim0.5$, we could not observe any sign of clear instability toward
either Neel or stripe type, at least for $T > 0.01$; the system is strongly correlated but remains to be
critical under the frustration. 

Let us compare the result to the $T=0$ argument by Moessner and Sondhi~\cite{Moessner2001}. In the previous
study, by a perturbation in $\Gamma/h$, it was predicted
that a Neel-type order occurs in the entire region of our consideration. Our result in Fig.~\ref{fig5}
shows a similar tendency in the intermediate $h$, where the Neel-type order appears to be favored.
However, we could not detect any sign of a particular ground state in the lower-$h$ region, although the Neel-type order
was expected to extend down to $h \to 0$ in the schematic $T=0$ phase diagram in the previous study~\cite{Moessner2001}.
This apparent disagreement can be due to very weak degeneracy lifting 
by the order from disorder for the expected Neel ground state.  
Another possibility is that,
for large $\Gamma/h$, the perturbative argument in $\Gamma/h$ no longer holds and a different phase emerges in the region.
In order to identify the nature of the system in this region, further analysis down to lower $T$ in larger
system sizes is necessary.
On the other hand, under relatively strong field $1.0 \lesssim h \lesssim 1.8$, 
we detected instability toward another phase 
from the slow relaxation behavior, namely, the horizontal-stripe phase.
This might be related with the competition in the ground states between 
the two-up two-down Neel order and 
three-up one-down diagonal-stripe order at $h \simeq 2$.
Our preliminary calculations show a finite-size effect in this regime; the
horizontal-stripe type instability appears to extend to slightly lower-$h$ region in larger size systems. More
detailed results will be discussed elsewhere. 

\begin{figure}[h]
\includegraphics[width=12pc]{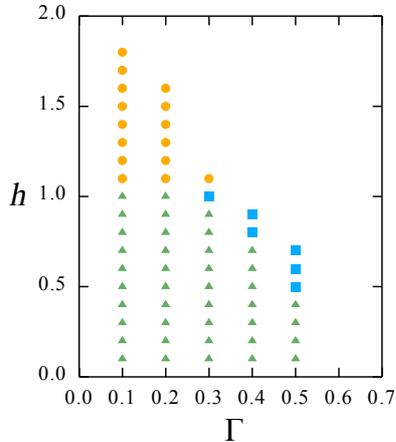}\hspace{2pc}%
\begin{minipage}[b]{20pc}
      \caption{
      `Phase diagram' for the model in Eq.~(\ref{eq:H}) as functions of  transverse field $\Gamma$ and
      longitudinal field $h$. The results indicate the low-$T$ instabilities which are deduced from the QMC
      results down to $T \simeq 0.01$. The blue squares (orange circles) show the region in which the
      Neel(horizontal stripe)-type instability shows up. The green triangles show the region where both
      $\chi_{\rm Neel}$ and $\chi_{\rm stripe}$ remain to show the Curie-law like divergence down to the
      lowest $T$. The area without symbols for $\Gamma>0$ indicates the quantum paramagnetic region.
      }
      \label{fig5}
\end{minipage}
\end{figure}

\section{Summary}

To summarize, we have investigated the effect of quantum and thermal fluctuations on the thermodynamics of
the frustrated checkerboard Ising model with transverse and longitudinal magnetic fields. We have examined
the magnetic instabilities induced by order-from-disorder mechanism by employing a continuous-time quantum
Monte Carlo method with the loop-flip update and replica exchange algorithm. 
We identified 
characteristic behaviors, which were not inferred from the previous study 
for the quantum order from disorder in the ground state. 
One is the instability toward a
horizontal-stripe-type ordering in a relatively high longitudinal-field regime. 
This is distinguished from the diagonal stripe predicted for a higher field.
The other is no clear indication of dominant ordering
or fluctuation down to $T \simeq 0.01J$ in a relatively low-field regime;
the susceptibility shows a persistent Curie-law-like divergence at
all the momenta. This could be due to the emergence of a novel state or simply by surprisingly weak
order-from-disorder effect in the Neel ground state expected in the previous theory.

H.I. and Y.M. thank T. Misawa, Y. Motoyama, H. Shinaoka, and M. Udagawa for helpful comments.
This research was supported by KAKENHI (No. 19052008 and No. 22540372), and Global COE Program``the Physical Sciences Frontier".

\end{document}